\documentstyle[aps,prl,epsfig,multicol]{revtex}
\relax   
\begin{document}
\preprint{prl}
\draft

\title{Quantum melting of the quasi-two-dimensional vortex lattice in $\kappa-$(ET)$_2$Cu(NCS)$_2$}

\author{M. M. Mola and S. Hill\cite{email}}
\address{Department of Physics, Montana State University, Bozeman, MT 59717}

\author{J. S. Brooks and J. S. Qualls}
\address{Department of Physics and National High Magnetic Field Laboratory, Florida State University,
Tallahassee, FL 32310}

\date{\today}
\maketitle

\begin{abstract}
We report torque magnetization measurements in regions of the
mixed state phase diagram (${\bf B}\sim\mu_o{\bf H}_{c2}$ and
T$_c$/$10^3$) of the organic superconductor
$\kappa-$(ET)$_2$Cu(NCS)$_2$ where quantum fluctuations are
expected to dominate thermal effects. Over most of the field range
below the irreversibility line (${\bf B}_{irr}$), magneto-thermal
instabilities are observed in the form of flux jumps. The abrupt
cessation of these instabilities just below ${\bf B}_{irr}$
indicates a quantum melting transition from a
quasi-two-dimensional vortex lattice phase to a quantum liquid
phase.
\end{abstract}

\pacs{PACS numbers: 74.70.Kn, 74.60.Ec, 74.60.Ge, 74.25.Ha}
%\bigskip

\begin{multicols}{2}[]
\sloppy
%\centerline{\bf Introduction}
%\smallskip
Although the properties of vortices in layered type$-$II
superconductors have been studied vigorously for over 15 years,
this subject continues to provide a rich and varied field for
investigation. The mixed superconducting state represents a
wonderful playground for studying general phase transformations
associated with vortex matter \cite{Blatter1}. In particular,
melting of the Abrikosov vortex lattice into a liquid phase has
drawn considerable interest, especially in extreme type$-$II
superconductors such as the high temperature superconductors
\cite{HTS} (HTS) and organic superconductors \cite{Inada}.
Moreover, identification of a melting transition driven by quantum
(as opposed to thermal) fluctuations has drawn much attention,
both theoretically \cite{quantummelt1} as well as experimentally
\cite{Blatter2}. A thorough understanding of vortex physics is
also essential for determining possible limiting behavior for
technological applications utilizing superconductors.

A model system for investigation of the structure and dynamics of
vortices in layered, type$-$II, superconductors is
$\kappa-$(ET)$_2$Cu(NCS)$_2$, where ET denotes
bis-ethylenedithio-tetrathiafulvalene \cite{Ishiguro}. Like the
HTS, this organic superconductor possesses a highly anisotropic
layered structure with the superconducting ET planes separated by
insulating anion layers \cite{Ishiguro}. The anisotropy parameter,
$\gamma$, defined as the ratio of the interlayer (currents
$\parallel$ {\em a}-axis) and in-plane (currents $\parallel$ {\em
bc}-plane) penetration depths ($\gamma \equiv
\lambda_a/\lambda_{bc}$), is thought to be in the range 50-200
\cite{gammaET}, {\em i.e.} similar to that of
Bi$_2$Sr$_2$CaCu$_2$O$_{8+d}$ \cite{gammabiscco}. Consequently, in
comparison to conventional superconductors, fluctuation effects
may be expected to play an important role at low-temperatures. In
contrast to the HTS, the organic superconductors are extremely
clean, with very few crystal defects. Furthermore, because of the
reduced T$_c$ and {\bf H}$_{c2}$ (T$_c = 9-10$ K
\cite{Ishiguro,lang,Zuo} and $\mu_o${\bf H}$_{c2} \approx 5$ T
($30-35$ T) \cite{lang,Zuo} for the field perpendicular (parallel)
to the superconducting layers), one is able to probe much more of
the temperature/field parameter space within the superconducting
state than is currently possible in the HTS. This, in turn, raises
the interesting prospect of studying two dimensional vortex
lattice melting over an extended temperature range (from T$_c$ to
T$_c$/10$^3$) $-$ possibly into the (low T,
high {\bf B}) quantum regime. %The parameter which defines the
%importance of quantum fluctuations, Q = R$_{eff}$/R$_Q$, where
%R$_{eff}$ = $\rho_n$/s, $\rho_n$ is the normal state resistivity,
%and s is the interlayer spacing, for this material is Q = 0.4.
%Such a large value (Q ~ 0.1 for cuprate superconductors) indicates
%the importance of quantum fluctuations at extremely low
%temperatures. Moreover,
Magnetic relaxation measurements performed by Mota {\em et al}.
\cite{Mota}, have shown that the crossover temperature from
thermal to quantum dominated fluctuations occurs at around 0.5 K
for this material.

%Consequently, we expect quantum effects to dominate the
%thermodynamics for our measurements, which were carried out
%between 25 and 200 mK.

It is well known from muon spin rotation ($\mu$SR) measurements
that a three-dimensional (3D) flux line lattice exists only at
very low fields ($<7 $mT$ \ll\mu_o${\bf H}$_{c2}$) in the title
compound \cite{Lee}; due to the large $\gamma$ value, the
individual quasi-two-dimensional (Q2D) vortex lattices in adjacent
layers become effectively decoupled above a roughly temperature
independent dimensional crossover field {\bf B}$_{cr}$ ($\sim
7$mT, see Fig. 1 and Ref. \cite{Lee}). Our recent temperature
dependent investigations of the interlayer Josephson plasma
resonance (JPR) in $\kappa-$(ET)$_2$Cu(NCS)$_2$ provide conclusive
proof that, in spite of the decoupling, long range Q2D order among
vortices within individual layers persists over most of the region
in the phase diagram between {\bf B}$_{cr}$ and the
irreversibility line (see shaded region in Fig. 1 and Refs
\cite{MolaJPR,Q2Dsolid}). Immediately below the irreversibility
line, local magnetization measurements \cite{Inada}, as well as
the JPR studies \cite{MolaJPR}, indicate a transition in this Q2D
vortex structure (see Fig. 1) $-$ Q2D melting has been put forward
as one possible explanation for this transition
\cite{Inada,MolaJPR}. These findings serve as a motivation for
investigating a wider range of the {\bf B},T phase diagram.
%Furthermore, recent reports of novel field-induced superconducting
%sub-phases in $\kappa-$(ET)$_2$Cu(NCS)$_2$, inferred from
%penetration depth measurements \cite{FFLO}, cry out for
%thermodynamic verification. Consequently, this letter reports
%angle dependent magnetic torque measurements at ultra-low
%temperatures and high magnetic fields.

High quality single crystals were grown using standard techniques
\cite{Ishiguro}. A single sample (approximate dimensions
$1.0\times1.0\times0.3$ mm$^3$) was mounted on a capacitive
cantilever beam torque magnetometer which, in turn, was attached
to a single axis rotator; $\theta = 0^o$ corresponds to the field
parallel to the least conducting {\em a}-axis, while $\theta =
90^o$ corresponds to the field parallel to the highly conducting
{\em bc}-planes. The sample, cantilever and rotator were then
loaded directly into the mixing chamber of a top-loading
$^3$He/$^4$He dilution refrigerator situated within the bore of a
20 T superconducting magnet at the National High Magnetic Field
Laboratory (NHMFL). For all measurements reported here, the
magnetic field was swept at a constant rate of 0.5 T/min.
Temperature dependent torque measurements were performed at
$\theta = 47^o$ and $74^o$. Subsequent analysis of the angle
dependence (from $\theta = 0^o$ to $90^o$) enabled us to scale the
temperature dependence back to $\theta = 0^o$, where the torque is
zero in this setup.

In Fig. 2, we plot the magnetization {\bf M}, derived directly
from the torque $\tau$ (${\bf M} = \tau/${\bf B}$\sin \theta$), as
a function of the applied magnetic field strength {\bf B}, for
angles between $\theta \sim 30^o$ and $79^o$ at approximately
$10^o$ intervals; the temperature is 25 mK. The overall shape of
these curves is consistent with previous measurements
\cite{Nishizaki,Sasaki}. The observed magnetization and associated
hysteresis (arrows in Fig. 2 indicate the field sweep direction)
are a consequence of the viscous flow of magnetic flux into (out
of) the sample upon increasing (decreasing) the applied magnetic
field strength. The hysteresis is greatest at low fields, and can
be seen to disappear completely above a characteristic field {\bf
B}$_{irr}$ ($< \mu_o{\bf H}_{c2}$), above which the sample behaves
reversibly. Perhaps the most pronounced features of the data in
Fig. 2 are the abrupt magnetization jumps. These "flux jumps" have
been observed previously in this and other materials
\cite{Sasaki,Swanson,Legrand}, and are due to an avalanche
behavior associated with the reorganization of magnetic flux as it
enters the sample; a systematic analysis of this phenomenon can be
found in Ref. \cite{Mola2}.

In the region between B$_{cr}$ and B$_{irr}$, crystal defects
collectively pin the Q2D vortex lattices in each layer. Thus, in a
field swept experiment, there is a build up of flux near the
sample surface. This creates a field gradient at the sample edge
together with an associated surface current given simply by
Maxwell's equation: $\nabla\times{\bf B} = \mu_o${\bf J}$_c$,
where {\bf J}$_c$ is the in-plane critical current density
\cite{Tilley,Bean}. At extremely low temperatures, a thermal
boundary (Kapitza) resistance \cite{Kapitza} isolates the sample
from the surrounding cryogen bath. Viscous transport of vortices
across the sample edge (where the surface screening currents flow)
causes local heating which, in turn, reduces the critical current
density, leading to additional heating and so on.. In the absence
of an effective thermal link to the surroundings, this can result
in runaway thermal instabilities which cause macroscopic regions
of the sample to become metallic \cite{Legrand,Mints}. When this
occurs, flux is able to flow rapidly into the crystal's interior,
thereby negating any/most of the surface currents. The crystal
then quickly cools and once again becomes superconducting with a
slightly different metastable vortex arrangement; the process then
starts anew. Indeed, Legrand {\em et al.} \cite{Legrand}, have
shown that, for a field swept experiment, YBa$_2$Cu$_3$O$_7$
samples experience large temperature spikes in conjunction with
discontinuities in the magnetization. These jumps are then
followed by a rapid relaxation back to the surrounding bath
temperature, consistent with the model above. An analysis of the
temperature dependence of the magnitudes of the observed
magnetization jumps ($\propto$T$^{3/2}$) in Fig. 2 supports this
picture $-$ see Ref. \cite{Mola2}. An important consequence of
this model is that the observation of flux jumps depends on the
stiffness of the vortex structure, {\em i.e}. jumps will only be
observed when collective vortex pinning is strong. These flux
jumps can thus be used as an indication of the existence of a
vortex solid of some sort.

In Fig. 3 we plot close-ups of the (raw) data in Fig. 2 in the
vicinity of ${\bf B}_{irr}$. Notice that, for small angles (Fig.
3a), there is a noticeable kink in the magnetization at an angle
dependent field, denoted {\bf B}$_m$ (indicated by arrows), just
slightly below ${\bf B}_{irr}$. Prior to this kink, the flux jumps
decay $-$ becoming negligibly small in the vicinity of the kink
$-$ while, above the kink, the magnetization decreases smoothly to
the reversible domain.
%Comparing up- and down-sweeps, there is a noticeable hysteresis,
%both in the magnitude and the position of the kink. The
%observation of this kink, together with the hysteresis, is
%suggestive of a first-order phase transition.
At larger angles (Fig. 3b), flux jumps persist up to the kink, but
never beyond. In fact, at the largest angles, the amplitudes of
the flux jumps are very large $-$ much larger than the kink
amplitude $-$ and, thus, the kink is not discernible. However, an
abrupt cessation of the flux jumps is instead observed at a field
whose angle dependence merges smoothly into the angle dependence
of the kink field {\bf B}$_m$ (see Fig. 4a). Thus, we assume that
the dramatic flux jump cessation, and the kink observed at smaller
angles, are related. The fact that the flux jumps cease at {\bf
B}$_m$ suggests that the stiffness of the vortex system changes
$-$ possibly due to a melting transition (see below). In what
follows, we identify {\bf B}$_m$ by the kink at low angles, and
the flux jump cessation at high angles.

In Fig. 4a, we have plotted {\bf B}$_m$ and {\bf B}$_{irr}$ vs.
$\theta$. The solid lines are fits to the data using the following
scaling law derived from 2D Ginzburg-Landau theory \cite{Zuo}:

%for an anisotropic Ginzburg-Landau (GL) theory derived by Blatter
%{\em et al}. \cite{Blatter1}, where

\bigskip

%\[
%{\rm B}_m ({\rm \theta ,T})\, = \,{\rm B}_m ({\rm 0,T}){\rm
%f}({\rm \theta }),
%\]

%\noindent{where}\hfill (1)

%\[
%{\rm f}(\theta )\, = \,\left[ {\alpha ^{-2} \sin ^2 \theta \, +
%\,\cos ^2 \theta } \right]^{{\raise0.7ex\hbox{${ - 1}$}
%\!\mathord{\left/
% {\vphantom {{ - 1} 2}}\right.\kern-\nulldelimiterspace}
%\!\lower0.7ex\hbox{$2$}}}.
%\]

% MathType!MTEF!2!1!+-
% feaaeaart1ev0aaatCvAUfeBSjuyZL2yd9gzLbvyNv2CaerbuLwBLn
% hiov2DGi1BTfMBaeXatLxBI9gBaerbd9wDYLwzYbItLDharqqtubsr
% 4rNCHbGeaGqiVu0Je9sqqrpepC0xbbL8F4rqqrFfpeea0xe9Lq-Jc9
% vqaqpepm0xbba9pwe9Q8fs0-yqaqpepae9pg0FirpepeKkFr0xfr-x
% fr-xb9adbaqaaeGaciGaaiaabeqaamaabaabaaGcbaWaaqWaaeaada
% WcaaqaaGqabiaa-jeadaWgaaWcbaGaamyBaaqabaGccaGGOaGaaeiU
% diaacMcaciGGZbGaaiyAaiaac6gacaqG4oaabaGaa8NqamaaBaaale
% aacaWGTbGaeyyPI4fabeaaaaaakiaawEa7caGLiWoacaaMc8Uaey4k
% aSIaaGPaVpaabmaabaWaaSaaaeaacaWFcbWaaSbaaSqaaiaad2gaae
% qaaOGaaiikaiaabI7acaGGPaGaci4yaiaac+gacaGGZbGaaeiUdaqa
% aiaa-jeadaWgaaWcbaGaamyBaiablwIiqbqabaaaaaGccaGLOaGaay
% zkaaWaaWbaaSqabeaacaaIYaaaaOGaaGPaVlabg2da9iaaykW7caaI
% Xaaaaa!5C5A!
\[
\left| {\frac{{{\bf B}_c ({\rm \theta })\sin {\rm \theta }}}{{{\bf
B}_{c \bot } }}} \right|\, + \,\left( {\frac{{{\bf B}_c ({\rm
\theta })\cos {\rm \theta }}}{{{\bf B}_{c\parallel } }}} \right)^2
\, = \,1,
\]

%\[
%{\rm f}({\rm \theta })\, = \,\left[ {{\rm \gamma }^{\rm 2} \sin ^2
%{\rm \theta }\,{\rm  + }\,\cos ^2 {\rm \theta }} \right]^{ -
%{\raise0.3ex\hbox{$1$} \!\mathord{\left/
% {\vphantom {1 2}}\right.\kern-\nulldelimiterspace}
%\!\lower0.3ex\hbox{$2$}}}.
%\]

\bigskip

\noindent{where {\bf B}$_c$ refers to either {\bf B}$_m$ or {\bf
B}$_{irr}$, and the subscripts $\parallel$ and $\perp$ refer to
the limiting values of these fields with $\theta=90^o$ and
$\theta=0^o$, respectively; the same angle dependence has also
been noted for {\bf H}$_{c2}$ \cite{Zuo}. From the fit to ${\bf
B}_m(\theta)$, we obtain values for {\bf B}$_{m\perp}$ = 3.6 T and
{\bf B}$_{m\parallel}$ = 32 T, at T = 25 mK. Thus, ${\bf
B}_m(\theta)$ falls below {\bf B}$_{irr}$, and well below
$\mu_o{\bf H}_{c2}$ $-$ a region in {\bf B},T parameter space
where no such transition has previously been observed (earlier
low-temperature experiments measured only {\bf B}$_{irr}$
\cite{Sasaki}). Using two separate models
\cite{quantummelt1,Blatter2}, Sasaki {\em et al.} \cite{Sasaki},
have made a rough estimate of the T = 0 K, $\theta=0^o$ quantum
melting field {\bf B}$_{qm\perp}$ for
$\kappa-$(ET)$_2$Cu(NCS)$_2$. For either model, they obtain {\bf
B}$_{qm\perp}\sim 4$ T, very close to the transition field {\bf
B}$_{m\perp}$ we observe at 25mK. Thus, given the extremely low
temperature of these measurements (T$\sim$T$_c/10^3$), along with
the proximity to the expected T = 0 K quantum melting field, we
propose that the observed transition at {\bf B}$_m$($\theta$, $T <
200$mK) does in fact correspond to a quantum melting transition
between a Q2D vortex lattice phase and a quantum liquid phase. The
existence of a liquid phase dominated by quantum fluctuation
effects in the region between ${\bf B}_{irr}$ and $\mu_o{\bf
H}_{c2}$ has, indeed, been noted by other authors
\cite{Sasaki,Ito}}.

%Given this identification, the value obtained for {\bf
%B}$_{m\parallel}$ is interesting due to its proximity to
%$\mu_o${\bf H}$_{c2}$. This raises the possibility that there is
%no flux lattice melting transition for $\theta=0^o$ at T = 0 K,
%but rather a quantum critical point. Clearly this will be the
%focus of further investigations.

%We now turn to the parameter $\alpha$ which, according to
%Ginzburg-Landau theory, should be equivalent to the anisotropy
%parameter $\gamma$ \cite{Blatter1}. Clearly, this is not the case
%here. Nevertheless, our value for $\alpha$, obtained both from
%fits to ${\bf B}_m(\theta)$ and ${\bf B}_{irr}(\theta)$, is very
%close to the ratio of the parallel and perpendicular critical
%fields (${\bf H}_{c2\parallel}/{\bf H}_{c2\perp}\sim 7$
%\cite{Nam}), as has been noted by other authors for this material
%\cite{Nam,Ito}. Thus, most of the angle dependent properties of
%this superconductor appear to be well described in terms of the
%above scaling law.

%Although the GL theory gives excellent agreement with the data, we
%obtain a value for $\gamma$ that is rather low compared to the
%accepted value of 50$-$200 \cite{gammaET}. As of now, the
%disparity between our result and those of previous measurements is
%not clear. However, we note that the angle dependence of H$_{c2}$
%gives rise to a similar value for $\gamma$, using the same scaling
%law \cite{Nam}. Most likely, the GL model used here is not the
%appropriate theory for this material.

The temperature dependence of {\bf B}$_m$ provides further
evidence that quantum effects become important as T$\rightarrow$0
K. After scaling measurements made at $\theta = 47^o$ back to
$\theta = 0^o$, we plot {\bf B}$_m$ and ${\bf B}_{irr}$ as a
function of temperature in Fig. 4b. In this {\bf B},T regime, the
irreversibility line shows a linear temperature dependence,
consistent with previous measurements \cite{Sasaki}, while {\bf
B}$_m$ exhibits a definite negative curvature $[${\bf
B}$_m\propto(T_c-$T)$^\alpha$, $\alpha<1]$; no classical theory of
melting can account for this trend. Nevertheless, Blatter {\em et
al}. have observed a similar behavior in 2D superconducting films,
which they attribute to a crossover from quantum melting to a
thermally assisted dislocation mediated form of melting at higher
temperatures \cite{Blatter2}. The crossover temperature predicted
by Blatter {\em et al}. occurs at around T $\sim$ T$_c$/100, which
is precisely where we observe the negative curvature (or
crossover) in the temperature dependence of {\bf B}$_m$. Clearly,
measurements spanning a wider temperature range will be necessary
in order to elucidate the true nature of the melting at higher
temperatures. However, the fact that the phase boundary curves
directly towards the T = 0 K axis, as T$\rightarrow$0 K, is
strongly suggestive of a quantum phase transition.

%Thus, given the cessation of flux jumps at the transition, the
%extremely low temperature and high field of the measurements, and
%the proximity to the expected T = 0 K quantum melting field, we
%propose that the observed transition at {\bf B}$_m$($\theta$, T
%$<$ 100 mK) does in fact correspond to a quantum melting of the
%Q2D vortex lattice.

Measurements on a second larger sample produced similar flux jumps
\cite{excuse}, and exactly the same abrupt flux jump cessation
just below {\bf B}$_{irr}$, as seen in Fig 3b. In fact, a data
point for this second sample falls precisely on the melting curve
in Fig 4a. Although the flux jumps are sample dependent
\cite{excuse}, their cessation appears to be sample independent,
indicating that the supposed melting transition is intrinsic, {\em
i.e.} {\bf B}$_m$(T,$\theta$) is unaffected by sample shape, size,
or details of the sample's surface.

%We note that surface effects can be ruled out as the mechanism for
%the magnetization kink and flux jump cessation, as we have
%performed this experiment on multiple samples of different
%geometries, obtaining similar results in all cases. Plastic
%dislocations can also be ruled out as the underlying mechanism for
%the transition, due to the extreme purity and lack of crystal
%defects in the samples. The title compound is well known to be
%several orders of magnitude cleaner than the HTS, and recent
%simulations by Reichhardt {\em et al}. have shown that plastic
%flow of vortices will only occur in a strongly pinned sample.
%Comparing the average pinning potential depth measured by
%magnetization relaxation, U$_o$, to the natural energy scale of
%the system, $\varepsilon_o$ = $(\Phi_o)^2$/$(8\pi^2\lambda^2)$, we
%find U$_o/\epsilon_o$ = 2.1 x $10^{-7}$. For plastic dislocations
%to occur, the above ratio must be on the order of unity.

%We note that the true nature of the temperature dependence of {\bf
%B}$_m$ may be complicated by the temperature dependence of the
%thermal link between the sample and the cryogen bath, which
%increases with increasing temperature \cite{Kapitza}. Hence,
%further studies will be necessary to separate these effects and,
%thereby, determine the true temperature dependence of the melting
%transition. Nevertheless, the behavior seen in Fig. 4b is
%qualitatively similar to the quantum melting observed in
%superconducting films at very low temperatures \cite{Blatter2}.

The pronounced hysteresis in {\bf B}$_m$ could be taken as an
indication of $1^{st}$ order behavior, as could the kink in {\bf
M} observed at {\bf B}$_m$; note that, for several traces in Fig.
3a, the kink has the appearance of a discontinuous jump. However,
it is difficult truly judge the order of the transition based on
the hysteresis, given the many factors which contribute to
asymmetry in the {\bf M} vs. $\mu_o${\bf H} loops. Indeed, these
factors may lead to physically different transitions on up and
down sweeps. The approach from low fields is preceded by frequent
catastrophic $1^{st}$ order flux avalanches and temperature
spikes. Thus, disorder and/or dislocations could play a role in
the up-sweep melting, though the extreme quality of these samples
would seemingly rule out plastic or glassy behavior (our JPR
studies have shown the pinning in this material to be several
orders of magnitude weaker than in the HTS \cite{MolaJPR}). The
approach from high fields, on the other hand, simply involves a
transition from a weakly pinned liquid state \cite{Blatter1} to
the ordered state. These matters aside, we assert that it is the
abrupt cessation of flux jumps at {\bf B}$_m$, together with the
angle dependence of {\bf B}$_m$, that suggest a 2D melting
transition. From the temperature dependence of {\bf B}$_m$, we
additionally propose that the melting transition becomes $1^{st}$
order, and is driven by quantum fluctuations as T$\rightarrow 0$.

Taking this new high field/low temperature data, and combining it
with previous JPR \cite{MolaJPR}, local magnetization
\cite{Inada,Sasaki}, and $\mu$SR \cite {Lee} measurements, we can
construct a mixed state phase diagram for
$\kappa-$(ET)$_2$Cu(NCS)$_2$, as shown in Fig. 1. It is tempting
to connect the present data to the data from Refs. \cite{Inada}
and \cite{MolaJPR} with a smooth curve (see dashed line in Fig.
1). This would seem to suggest that both transitions correspond to
Q2D melting. We should note, however, that the high temperature
phase line may be due to a thermally assisted depinning
transition, as noted in Ref. \cite{MolaJPR}. Finally, the present
measurements have been unable to detect any indication of the
transition to a Fulde-Ferrell-Larkin-Ovchinnikov state reported
recently by Singleton {\em et al.} close to $\theta = 90^o$
\cite{FFLO}.

%Given the temperature dependence of the JPR measurements, these
%transitions must be due to changes in the Q2D vortex structure.

%Although the true temperature dependence of the high field melting
%transition is as yet unknown, we are confident in the position of
%the melting field at the lowest temperatures. Thus, we can connect
%through both the low and high temperature data,

%It is interesting to note that the transition occurs below the
%irreversibility line at high field, then seems to move
%continuously closer to the reversible domain as the temperature is
%increased, and the transition field decreases.

%Although flux jumps are still observed very close to $90^o$, we
%were not able to observe the melting transition, or to access the
%reversible domain, for $\theta$ exactly equal to $90^o$ due to the
%field limit of the magnet used in this experiment.

%We note, however, that by extrapolating B$_m$ to $\theta=90^o$,
%using the scaling relation in Eq. 1 \cite{90degnote}, we find that
%the melting transition falls approximately at the same field as
%the structure interpreted as a transition to a The significance of
%this coincidence is, as yet, unclear. Clearly, further
%thermodynamic studies using the 33 T resistive magnets at the
%NHMFL, in conjunction with a dilution refrigerator, are needed to
%resolve this issue, and to follow the melting transition all the
%way out to $\theta=90^o$.

In conclusion, we have observed magneto-thermal instabilities in
the mixed state of the organic superconductor
$\kappa-$(ET)$_2$Cu(NCS)$_2$, which are associated with
transitions between metastable Q2D vortex lattice phases. The
abrupt cessation of magnetization jumps associated with these
instabilities serve as an indication of a melting of this Q2D
vortex lattice phase. Furthermore, this study $-$ which is the
first of its kind in the high-{\bf B}/low-T limit for such a
highly anisotropic superconductor $-$ suggests that the melting
may be driven by quantum rather than thermal fluctuations. Future
investigations will focus on the temperature dependence of this
transition, and on its evolution away from the high-{\bf B}/low-T
limit.

%Previous experiments have also observed a similar transition in a
%different region of this B-T phase space, and we believe these
%phenomenon may be related. Future studies will be directed towards
%filling in the intermediate field and temperature range, as well
%as determining the nature of the lattice melting in the high field
%regime.

This work was supported by the National Science Foundation
(DMR-0071953) and the Office of Naval Research (N00014-98-1-0538).
Work carried out at the NHMFL was supported by a cooperative
agreement between the State of Florida and the NSF under
DMR-9527035.

% References

%\clearpage

%\noindent{{\bf Figure captions}}

%\bigskip

%Fig. 1. {\bf B},T mixed state phase diagram for
%$\kappa-$(ET)$_2$Cu(NCS)$_2$, as deduced from JPR \cite{MolaJPR},
%$\mu$SR \cite{Lee} and magnetization measurements (this study and
%Refs \cite{Inada} and \cite{Sasaki}). Note that the Q2D solid
%(shaded region) and liquid phases occupy most of the available
%phase space, while the 3D solid phase exists only for very small
%applied fields ($<$7 mT).

%\bigskip

%Fig. 2. Magnetization as a function of magnetic field (up- and
%down-sweeps) for angles $\theta$ between $\sim30^o$ and $79^o$;
%the temperature is 25 mK. A pronounced hysteresis and "flux jumps"
%are clearly observed for all angles.

%\bigskip

%Fig. 3. Magnetization vs. magnetic field at low (a) and high (b)
%angles $\theta$ (indicated in the figure). In a), a distinct kink
%may be observed in the magnetization (indicated by arrows) at a
%field ${\bf B}_m$ just below ${\bf B}_{irr}$; the flux jumps decay
%smoothly to zero just before ${\bf B}_m$. At higher angles (b),
%the kink at ${\bf B}_m$ is obscured by large amplitude flux jumps.
%However, the abrupt cessation of these flux jumps serves as an
%alternative indication of the phase transition.

%\bigskip

%Fig. 4. (a) Angle dependence of the melting and irreversibility
%fields, ${\bf B}_m(\theta$) and ${\bf B}_{irr}(\theta$)
%respectively; the solid lines are fits to the scaling law in Eq.
%1. (b) Temperature dependence  of the melting and irreversibility
%fields.

\end{multicols}

%\clearpage

\noindent{{\bf Figure captions}}

\bigskip

Fig. 1. The mixed state {\bf B},T phase diagram for
$\kappa-$(ET)$_2$Cu(NCS)$_2$. The legends correspond to:
$\diamondsuit$ $-$ $\mu_o${\bf H}$_{c2}$ [9]; $\nabla$ $-$ {\bf
B}$_{irr}$ [16]; $\bullet$ $-$ 2D melting or depinning [13];
$\Box$ $-$ first-order transition [3]; $\bigtriangleup$ $-$ 3D to
2D crossover [12]; $\circ$ $-$ 2D melting (this study). Note that
the Q2D solid (shaded region) and liquid phases occupy most of the
available {\bf B},T$-$space.

\bigskip

Fig. 2. Magnetization as a function of magnetic field (up- and
down-sweeps) for angles $\theta$ between $\sim30^o$ and $79^o$;
the temperature is 25 mK.

\bigskip

Fig. 3. Magnetization vs. magnetic field at (a) low and (b) high
angles $\theta$ (indicated in the figure). In (a), a distinct kink
may be observed in the magnetization (indicated by arrows) at a
field ${\bf B}_m$ just below ${\bf B}_{irr}$; the flux jumps decay
smoothly to zero just before ${\bf B}_m$. At higher angles (b),
the kink at ${\bf B}_m$ is obscured by large amplitude flux jumps.
However, the abrupt cessation of these flux jumps serves as an
alternative indication of the phase transition.

\bigskip

Fig. 4. (a) Angle dependence of the melting and irreversibility
fields, ${\bf B}_m(\theta$) and ${\bf B}_{irr}(\theta$)
respectively; the solid lines are fits to the scaling law in Eq.
1. (b) Temperature dependence  of the melting and irreversibility
fields.

%\end{multicols}

\clearpage

\begin{figure}
\centerline{\epsfig{figure=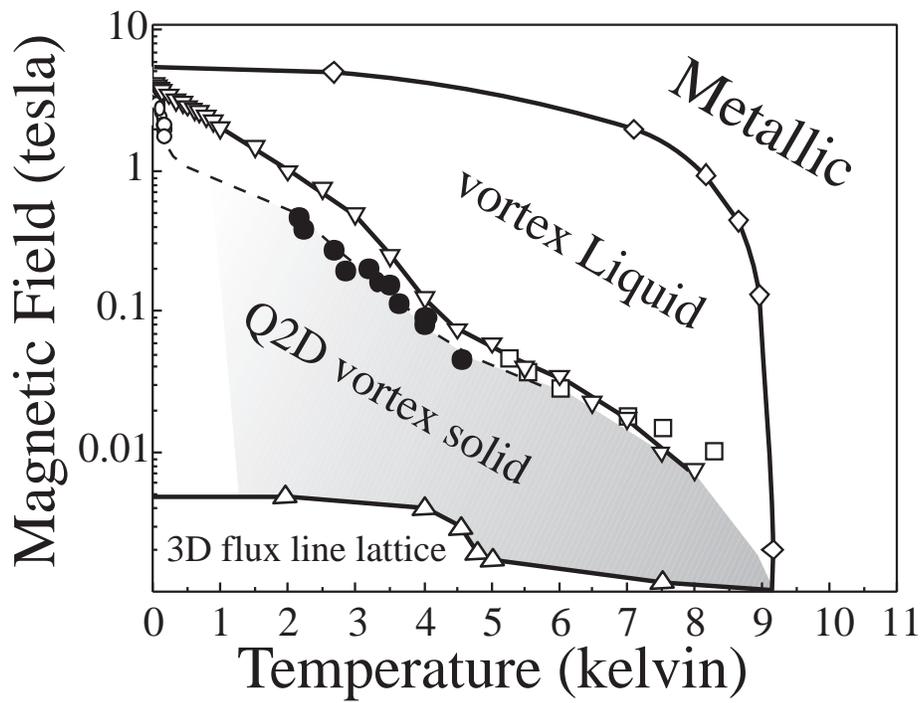,width=120mm}}
\bigskip
\caption{Mola {\em et al.}} \label{Fig. 1}
\end{figure}

\clearpage
%\clearpage

\begin{figure}
\centerline{\epsfig{figure=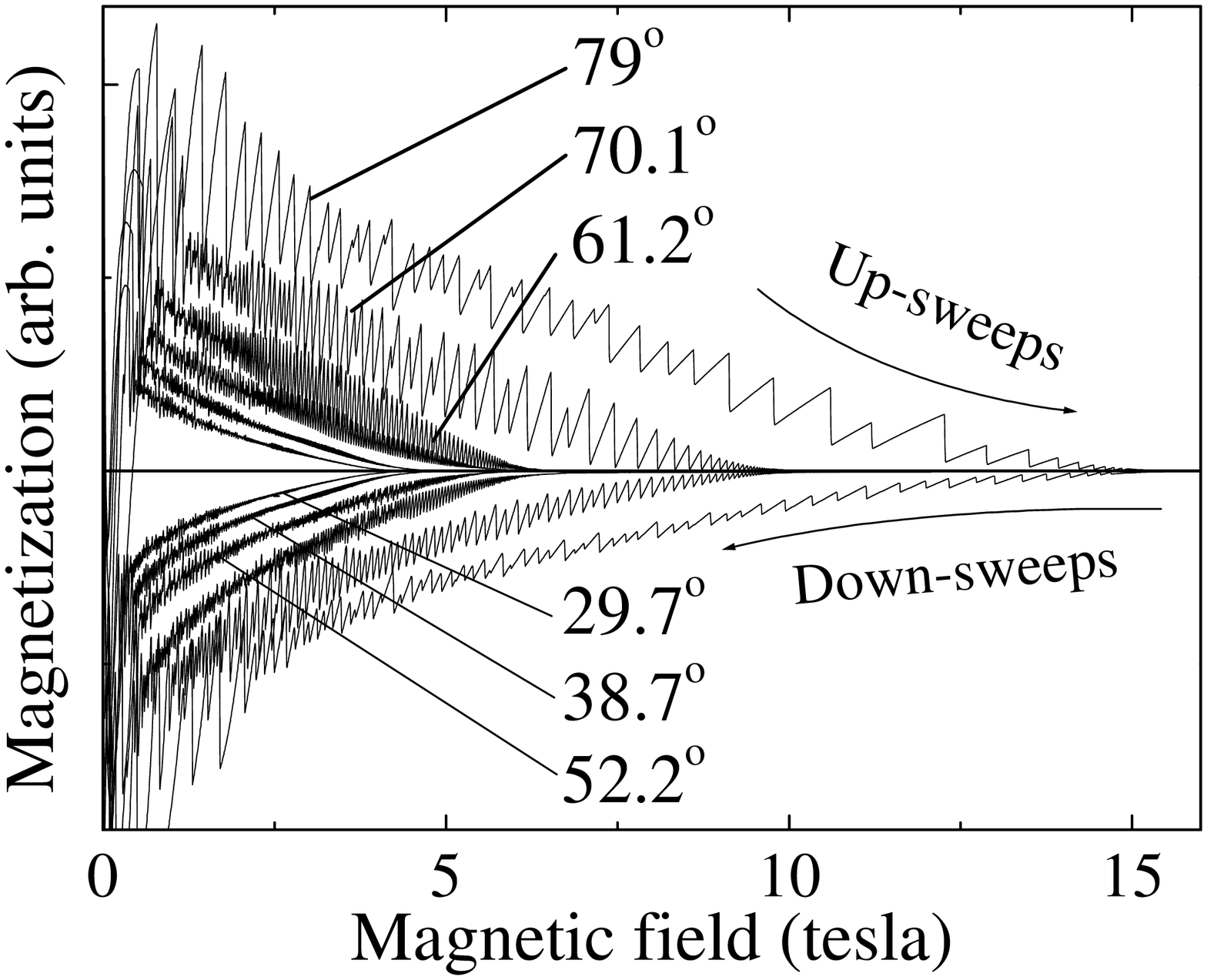,width=140mm}}
\bigskip
\caption{Mola {\em et al.}} \label{Fig. 2}
\end{figure}

\clearpage

\begin{figure}
\centerline{\epsfig{figure=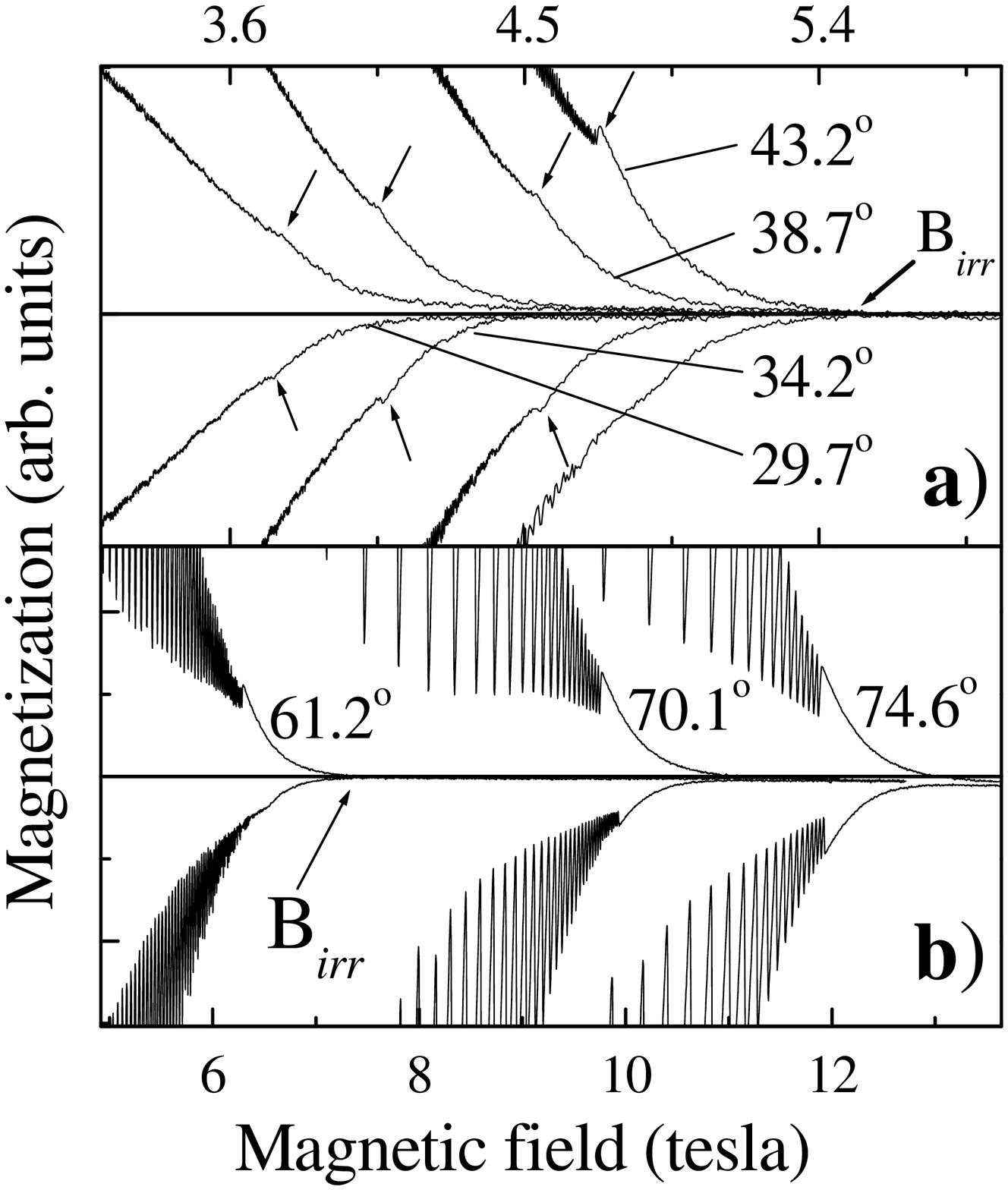,width=120mm}}
\bigskip
\caption{Mola {\em et al.}} \label{Fig. 3}
\end{figure}

\clearpage
%\clearpage

\begin{figure}
\centerline{\epsfig{figure=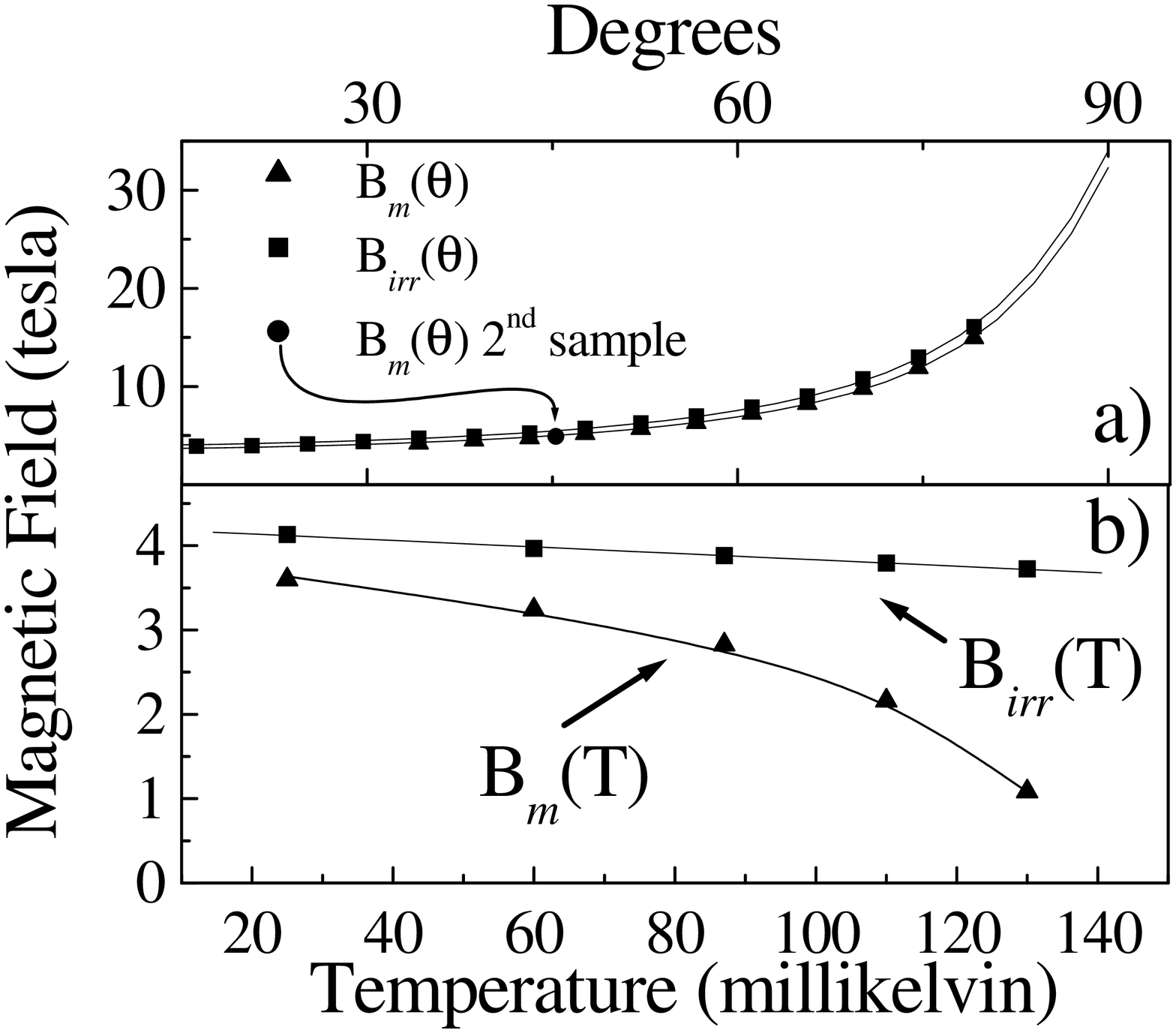,width=120mm}}
\bigskip
\caption{Mola {\em et al.}} \label{Fig. 4}
\end{figure}

\bigskip

%\end{multicols}

\begin{references}

\bibitem[\dag]{email}
email: hill@physics.montana.edu

\bibitem{Blatter1}
G. Blatter {\em et al.}, Rev. Mod. Phys. {\bf 66}, 1125 (1995).

\bibitem{HTS}
See G. W. Crabtree and D. R. Nelson, Physics Today {\bf 50}, 38
(AIP, April 1997); and references therein.

%\bibitem{HTS2}
%A. Schilling {\em et al.}, Nature {\bf 382}, 791 (1996).

%\bibitem{OS1}
%T. Shibauchi {\em et al.}, Phys. Rev. B {\bf 57}, R5622 (1998).

\bibitem{Inada}
M. Inada {\em et al.}, J. Low Temp. Phys. {\bf 117}, 1423 (1999);
and references therein.

\bibitem{quantummelt1}
A. Rozhkov and D. Stroud, Phys. Rev. B {\bf 54}, R12697 (1996).

\bibitem{Blatter2}
G. Blatter {\em et al.}, Phys. Rev. B {\bf 50}, 13013 (1994).

\bibitem{Ishiguro}
T. Ishiguro, K. Yamaji, and G. Saito, {\em Organic
Superconductors} (Springer-Verlag, Berlin, 1998).

\bibitem{gammaET}
A. Carrington {\em et al.}, Phys. Rev. Lett. {\bf 83}, 4172
(1999); and references therein.

\bibitem{gammabiscco}
Y. Matsuda {\em et al.}, Phys. Rev. Lett. {\bf 78}, 1972 (1997).

\bibitem{lang}
M. Lang {\em et al}., Phys. Rev. B {\bf 49}, 15227 (1994).

\bibitem{Zuo}
F. Zuo {\em et al}., Phys. Rev. B {\bf 61}, 750 (2000); and
references therein.

\bibitem{Mota}
A. C. Mota {\em et al}., Physica C {\bf 185-189}, 343 (1991).

\bibitem{Lee}
S. L. Lee {\em et al.}, Phys. Rev. Lett. {\bf 79}, 1563 (1997); F.
L. Pratt {\em et al.}, submitted.

\bibitem{MolaJPR}
M. M. Mola {\em et al.}, Phys. Rev. B {\bf 62}, (2000).

\bibitem{Q2Dsolid}
The only theoretical models that can account for the temperature
dependence observed in Ref. \cite{MolaJPR} assume pinned Q2D solid
vortex lattices in each layer, with no correlation between the
locations of vortices in adjacent layers.

\bibitem{Nishizaki}
T. Nishizaki {\em et al.}, Phys. Rev. B {\bf 54}, R3760 (1996).

\bibitem{Sasaki}
T. Sasaki {\em et al.}, Phys. Rev. B {\bf 57}, 10889 (1998).

\bibitem{Swanson}
A. G. Swanson {\em et al.}, Solid State Comm. {\bf 73}, 353
(1990).

\bibitem{Legrand}
L. Legrand {\em et al.}, Physica C {\bf 211}, 239 (1993).

\bibitem{Mola2}
M. Mola {\em et al}., cond-mat/0011227.

\bibitem{Tilley}
D. R. Tilley and J. Tilley, {\em Superfluidity and
Superconductivity} (IOP Publishing, London, 1990).

\bibitem{Bean}
C. P. Bean, Rev. Mod. Phys. {\bf 36}, 31 (1964).

\bibitem{Kapitza}
A. Kent, {\em Experimental Low-Temperature Physics} (AIP, New
York, 1993).

\bibitem{Mints}
R. G. Mints and A. L. Rakhmanov, Rev. Mod. Phys. {\bf 53}, 551
(1981).

%, J. Phys. D: Appl. Phys. 12 (1979) 1929.


%\bibitem{Nam}
%M-S. Nam {\em et al.}, J. Phys.: Cond. Mat. {\bf 11}, L477 (1999).

\bibitem{Ito}
H. Ito {\em et al.}, J. Superconductivity {\bf 12}, 525 (1999).

\bibitem{excuse} Limited data were obtained for this second sample
due to the fact that its larger magnetic moment overwhelmed the
detection scheme of the magnetometer.

\bibitem{FFLO}
J. Singleton {\em et al.}, J. Phys.: Condens. Matter {\bf 12},
L641 (2000).

\end{references}
\end{document}